\documentclass{PoS}

\title{Low Frequency GMRT Observations of Microquasar V4641 Sgr}

\ShortTitle{Low Frequency GMRT Observations of V4641 Sgr}

\author{\speaker{C. H. Ishwara-Chandra}, Sabyasachi Pal and A. Pramesh Rao\\
        National Centre for Radio Astrophysics, Pune, India\\
        E-mail: \email{ishwar@ncra.tifr.res.in, spal@ncra.tifr.res.in, pramesh@ncra.tifr.res.in}}

\abstract{We have observed the micro-quasar V4641 Sgr (SAX J1819.3-2525) 
at the time of flare during May 2002, using the Giant Metrewave Radio Telescope at
low radio frequencies (244 and 610 MHz). This is the lowest
frequency radio detection of the source. The source showed
clear signature of spectral evolution from optically thick
to thin state.  We model the spectral evolution of the source to obtain
some of the source parameters. Assuming the value of magnetic field strength to be in the range of 0.1 to 1 Gauss, we obtain the 
initial velocity of expansion of the jet to be in the range $\sim$ 0.45c to 0.25c. 
This is consistent with the known 
expansion velocities of the jet with other microquasars.

}

\FullConference{VI Microquasar Workshop: Microquasars and Beyond\\
		September 18-22 2006\\
		Societ\`a del Casino, Como, Italy}

\begin{document}
\section{Introduction}
The micro-quasar V4641 Sgr (XTE J1819-254/SAX J1819.3-2525) was discovered
independently by proportional counter array in Rossi X-ray Timing
Explorer (RXTE) \cite{Ma99} and wide field cameras on board BeppoSAX
\cite{Za99} in 1999. Optical spectroscopy and photometry showed that
V4641 Sgr is a black hole candidate with a mass of $\sim 9.6 M_\odot$ in a
binary system with a high mass companion \cite{or01}. The radio morphology
of the source suggests that it is a relativistic jet source like SS433,
GRS 1915+105 and Sco X-1 \cite{Hj00} and thus it is a microquasar. The
object showed flaring events in past in X-ray \cite{Ma06}, optical
\cite{Ue02} and radio \cite{Hj00} wavelengths. 
From RXTE-PCA monitoring of galactic centre region
on UT 17.6 and 20.9 May 2002, Markwardt and Swank \cite{Ma02} reported
that the source had started flaring. Here we report low
frequency radio observation of the source in 244 and 610 MHz using the Giant
Meterwave Radio Telescope (GMRT) during the May 2002 outburst. In Section 2,
we briefly describe the observation and data analysis 
and in Section 3 we discuss the results. 
The summary and conclusions are presented in Section 4. 

\begin{table*}[b]
 \centering
 \begin{minipage}{140mm}
  \caption{Observation Log and Fluxes}
  \begin{tabular}{@{}cccccc@{}}
  \hline
Date 2002 &  MJD &Frequency& Duration & Flux (mJy) &Flux (mJy)\\
May (UT)&  &(MHz)&(sec)&V4641 Sgr&background\\
&&&&&source\\
 \hline
23.99  &52417.99  &244  & 1680 &$57.8\pm7.70$&$102.9\pm6.3$\\
24.89  &52418.89  &244  &2700  &$81.8\pm5.04$&$105.0\pm8.2$\\
27.85  &52421.85  &244  & 1800 &$\S<19.9$&$105.8\pm10.1$\\
28.90  &52422.90  &244  & 1800 &$\S<20.3$&$107.5\pm13.6$\\
       &&&&&\\
24.03  &52418.03  &610  &1920&$101.8\pm5.25$  &$65.2\pm3.39$\\
25.02  & 52419.02 &610  &  1500&$102.5\pm5.33$&$66.3\pm3.44$\\
25.92  & 52419.92 &610  &  4500&$47.8\pm2.97$&$71.0\pm3.66$\\
26.92  & 52420.92 &610  &  4800&$14.4\pm1.75$&$71.3\pm3.68$\\
28.01  & 52422.01 &610  &  2700&$7.6\pm1.38$&$67.9\pm3.51$\\
29.76  & 52423.76 &610  & 1680&$1.4\pm1.28$ &$62.8\pm3.20$\\
\hline
\label{tab1}
\end{tabular}

{$\S$ $5\sigma$ upper limit}
\end{minipage}
\end{table*}

\section{Observation and Data Analysis}

The present observation of V4641 Sgr is carried out at low radio frequencies of 244 and 610 MHz
using the Giant Metrewave Radio Telescope (GMRT) \cite{sw1} during the radio
flare in May 2002. GMRT consists of 30 fully steerable parabolic antennas
out of which 16 are distributed in a nearly `Y' shaped array and the remaining
14 antennas are randomly distributed in the central 1 km region. The resolution
at 244 and 610 MHz are $13^{\prime\prime}$ and $5^{\prime\prime}$
respectively. Details of GMRT specification can be found at GMRT home
page {\tt www.gmrt.ncra.tifr.res.in}. The bandwidths of the present
observations were
6 MHz and 16 MHz at 244 and 610 MHz respectively in the spectral line
mode with a total of 128 channels with channel width of 125 kHz per
channel. For our observations, we used 16.9 second integration
time. Presently, GMRT has a facility to observe simultaneously in
244 and 610 MHz but this facility was not available at the time of
this observation. However, whenever possible, we have taken near simultaneous
observation at 244 and 610 MHz. We have used 3C48 or 3C286 as flux and
bandpass calibrator and 1626-298 as phase calibrator. The observation
log and the measured flux of V4641 Sgr on different days are presented in
Table \ref{tab1}.
The flux of a background source J181929$-$253736 is also included
in the Table, which exhibits constant flux.

We have also used the archival data of Very Large Array (VLA), observed
between 23rd May and 31st May 2002 at 1.5, 4.9, 8.4, 15.0, 22.5 and 43.3
GHz. The array was in the AB
configuration at that time. The data obtained from VLA is processed
by using AIPS with standard procedures.

The data recorded with GMRT was converted to FITS format and then analyzed
with the Astronomical Image Processing System (AIPS) using standard procedures. 
After editing the data, the data in 128 channels was collapsed into fewer channels by 
applying bandpass. To take care of bandwidth
smearing, at 244 MHz, we have averaged each 8 channels. The flux measurements at 
244 and 610 MHz were corrected for the increased sky background in the
direction of V4641 Sgr. 

\begin{figure}
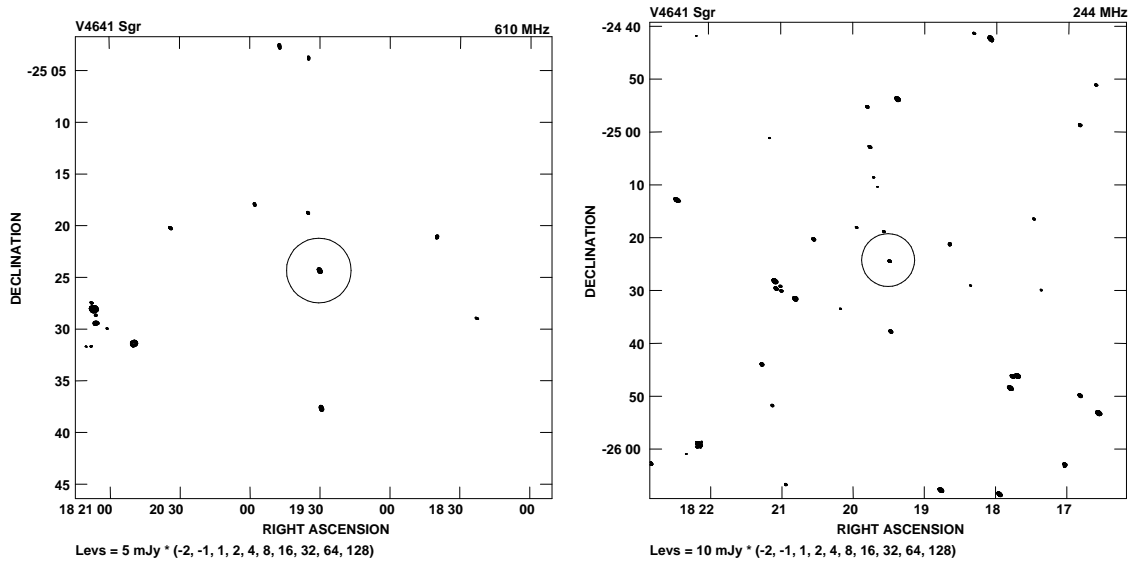

\includegraphics[width=0.5\textwidth]{chandra_v46_f1.ps}
\includegraphics[width=0.5\textwidth]{chandra_v46_f2.ps}
\caption{GMRT images of V4641 Sgr in 610 MHz (left) and 244 MHz (right).
The circle drawn at the central region is only to help to
locate the V4641 Sgr, which is at the center of this circle.
}
\label{image_244_610}
\end{figure}

\section{Results and Discussion}

\subsection{Radio Light Curve and Spectral Evolution}

The image of the field of V4641 Sgr at 244 and 610 MHz is presented 
in Figure \ref{image_244_610}. The source V4641 Sgr is indicated by a
circle as on 24 May at 610 MHz and 244 MHz.
The source is not resolved at any of these frequencies. There
are many background sources in the field. The flux of these background
sources is constant in all the days of observations within measurement errors. 
V4641 Sgr is clearly detected at 610 and 244 MHz over the first two
days. While the flux remained steady at 610 MHz on May 24 and 25, there
was an increase in flux at 244 MHz over the same duration, indicating that the
spectra has become optically thin on the second day. 
The radio light curve is presented in Figure 2. The source is seen to fade
from 25 to 30 May 2002. The flux density is below $5\sigma$
limit of $\sim$ 20 mJy on UT 27.85 May 2002 in 244 MHz, though it was still visible in 610 MHz
with flux density of 7.6mJy (averaged over whole scan). The $5\sigma$ limit at 244 MHz 
is significantly higher than the detected flux at 610 MHz, 
hence it is possible that the source is optically thin. 
In the Table
\ref{tab1} we have listed the flux of the source (averaged over whole
scan) for different days in each frequencies and for comparison the
flux density of a background source J181929$-$253736 is also given. The
radio data at 610 MHz is roughly fitted with an empirical decay formula
$\sim 0.92 \nu^{0.51}_{MHz} (MJD-52419)^{-0.37}$. In Figure 2 (bottom panel), we have
plotted the variation of spectral index with time between 244 and 610 MHz,
assuming $S_\nu\propto{\nu}^{\alpha}$. The plot clearly shows spectral
evolution from optically thick to optically thin states. The spectral
index $\alpha$ on May 24, 2002 was $\sim$ 0.60 which changed to 0.24 on next
day. The source showed high variability for each days of observation. In
Figure \ref{dftpl}, we show the variability of the source for the
observation of four days at 610 MHz. No signature of any periodicity or
quasi periodic oscillation is found from the fast Fourier transform of
short scans of variable radio data.

\begin{figure}
\includegraphics[width=.8\textwidth,angle=-90]{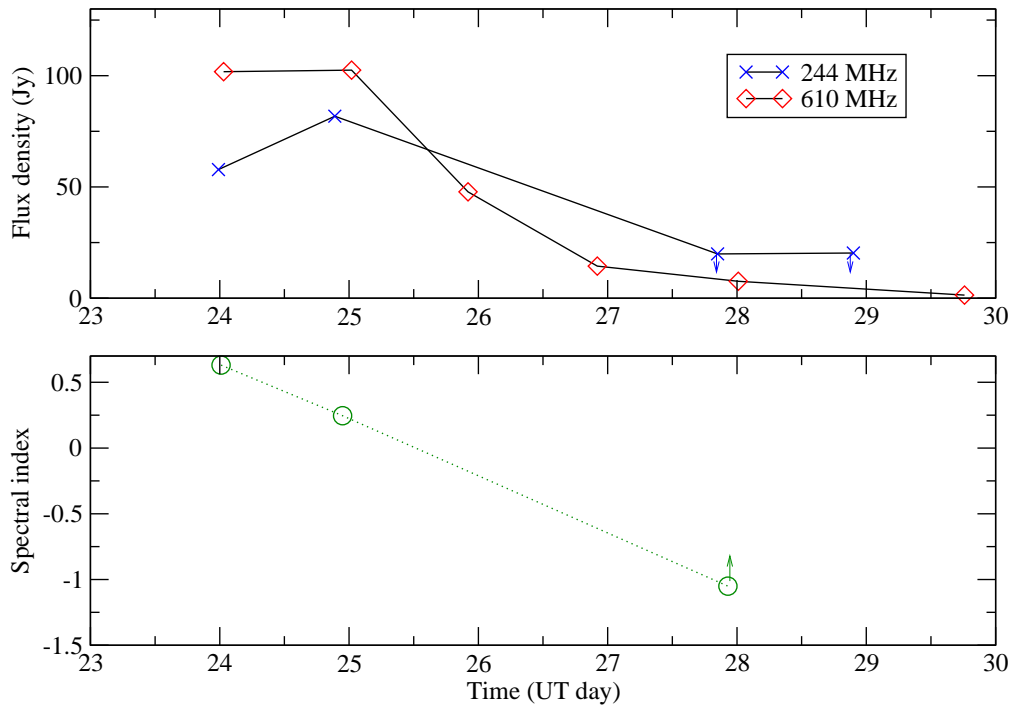}
\caption{In the top panel, the radio light curve at 610 and 244 MHz is
presented. The bottom panel shows the spectral index for the days where
near simultaneous data was available between 610 and 244 MHz.}
\end{figure}

\begin{figure}
\includegraphics[width=.8\textwidth,angle=-90]{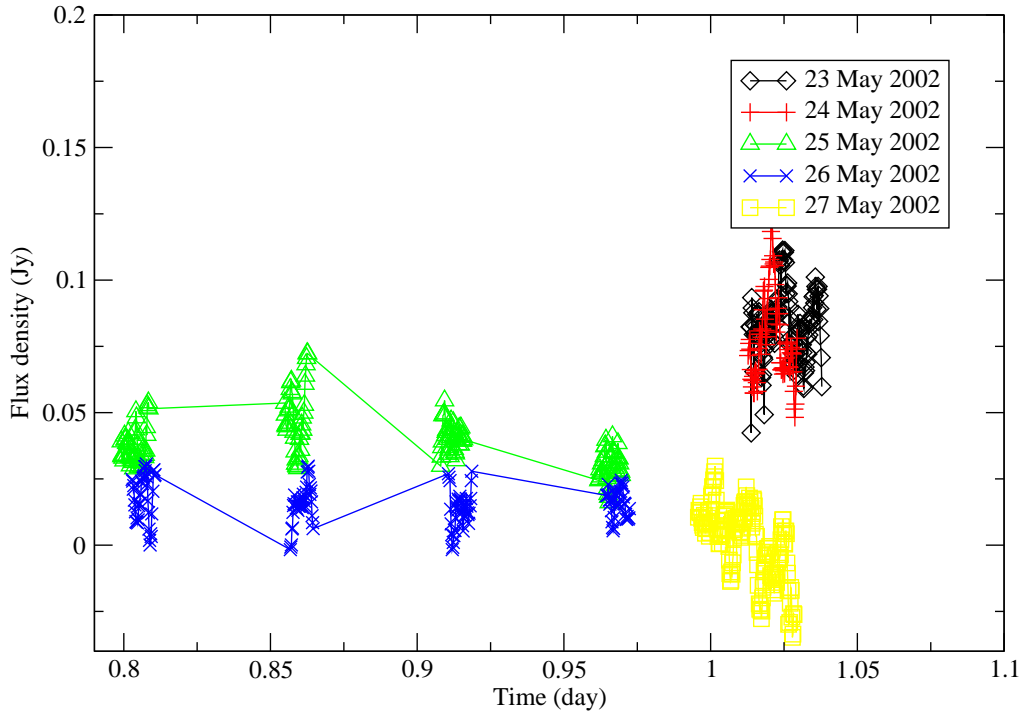}
\caption{Variation of radio flux density at 610 MHz obtained on 24, 25, 26, 27 and 28 May 2002. 
Data is obtained using the Giant Meterwave Radio Telescope (GMRT).} 
\label{dftpl}
\end{figure}

\subsection{Jet Expansion Speed}

If we assume the absorption in the optical thick region is due to
synchrotron self absorption, it is possible to calculate the quantity
$\theta/B^{1/4}$ (containing the size of the radio source $\theta$
and the magnetic field $B$) by equating the the observed flux with the
source function for the non thermal emission (as done for the case of
SS433 in Pal et al \cite{Sa06}). From the standard synchrotron synchrotron self-absorption formulae,
\cite{Mo72}

\begin{equation}
\theta^2 \approx 900S_s^*B^{1/2}F'\nu_s^{-2.5}
\end{equation}

where 
$\theta$ is in seconds of arc, 
$\nu_s$ is the turnover frequency in MHz,
$S_s^*$ is the flux density at the turnover frequency in Jy,
$F'$ is a function of the power-law index $p ~ (= 2\alpha + 1)$
and $B$ is the magnetic field in Gauss.

\begin{figure}
\includegraphics[width=.68\textwidth,angle=270]{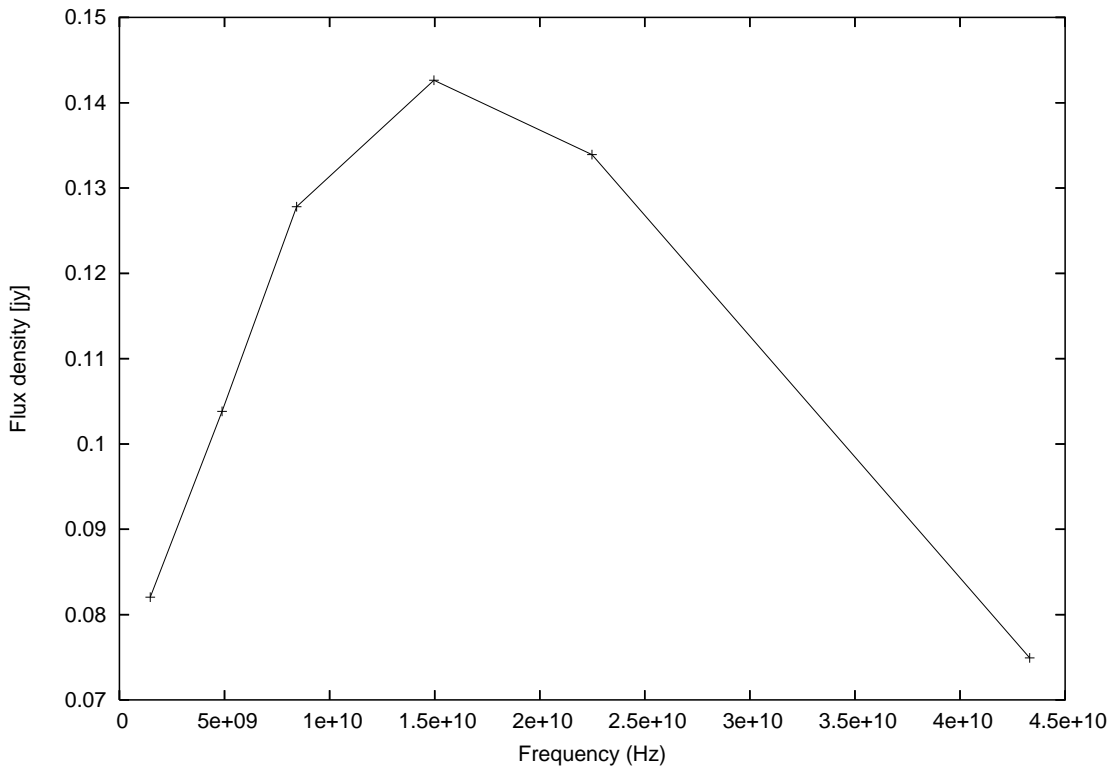}
\caption{Radio spectrum of V4641 Sgr on 23rd May 2002 from VLA archival data.} 
\label{vla_spectrum}
\end{figure}

From Figure \ref{vla_spectrum}, where we have shown the radio
spectrum of the source on 23rd May 2002 (UT 08h to 12h), the turnover frequency is approximately
$\nu_s\sim 17$ GHz and the
spectral index $\alpha$ between 22.5 and 43.3 GHz is -0.89 (assuming
$S_\nu \propto \nu^{\alpha}$). This implies $\theta/ B^{1/4}\sim 0.13$, 
where $\theta$ is in milli-arcsec and B in Gauss. If we assume
$B\sim100$ mG, the size of the radio source will be $\theta \sim
0.23$ mas.

The source has expanded since then and the spectra has become optically thin. 
The source is not detected in further VLA observations at GHz frequencies a few days later.
From the GMRT light curve, the synchrotron turn over frequency did not
cross below 610 MHz till 25th May.  Unfortunately there
is no measurement at 244 MHz on this day or on 26th May, but the examination of the light
curve at 244 MHz suggests that the peak should have occurred after 25th May, but before 27th May.
The time at which the 244 MHz emission would have peaked can
be estimated using the adiabatic expansion models
used to estimate the time delay for GRS1915+105 \cite{Ic02}.
Assuming an optically thin spectral index of $\sim$ 0.75, the application of the 
above model gives us a time delay of $\sim$ 2.7 days between the peak at 17 GHz and 244 MHz.
This means the 244 MHz emission would have peaked on May 26.1. 
This is broadly consistent with the light curve, where the peak at 244 MHz was expected after 25th May, but before 27th May, hence we use this time for 
further calculations. From the synchrotron self absorption equation, $\theta/ B^{1/4} \sim 20$
if the turnover frequency and flux is 244 MHz and 100 mJy respectively.
Assuming a magnetic field $B\sim100$ mG, we get the size of the
radio source to be $\theta \sim 35.7$ mas.

It should be noted that the size of the source depends weakly on
the value of the magnetic field $B$. From the expansion between the days mentioned above, 
we can obtain the source expansion speed. 
Taking the distance of V4641 Sgr to be 6.1 kpc \cite{Or00}, the velocity
of expansion comes out to be $\sim 0.45c$.
If we assume the magnetic field to be 1 Gauss, the expansion speed is $\sim$ 0.25c.

\section{Conclusion}
We have shown for the first time that V4641 Sgr is active at meter
wavelengths. This is the first detection of the source at the low
frequency of 244 MHz. The source has shown considerable variability at 610 MHz.
Synchrotron self absorption explains the
behavior of the source in the optically thin limit with reasonable extent.
Assuming the radio source is expanding spherically, we get the velocity of expansion 
of $\sim 0.45$c to 0.25c for magnetic field values of 0.1 to 1 Gauss. This is broadly in agreement with the expansion speeds observed in some of the microquasars.

\label{}

\section*{Acknowledgments}
We thank the staff of the GMRT that made these observations possible. GMRT
is run by the National Centre for Radio Astrophysics of the Tata Institute
of Fundamental Research. We have used data from National Radio Astronomy
Observatory (NRAO) archive. The NRAO is a facility of the National
Science Foundation operated under cooperative agreement by Associated
Universities, Inc.

\end{document}